# Pion Production Measurement in NA61/SHINE Experiment for High Precision Neutrino Oscillation Experiments

**Tomasz Jan Palczewski**[1]
*for NA61/SHINE Collaboration.*
*Soltan Institute for Nuclear Studies*
*Warsaw, Poland.*
*E-mail:* tomasz.palczewski@fuw.edu.pl

One of physics goals of the NA61/SHINE experiment is a measurement of hadron production cross sections from proton-Carbon interactions at 31GeV/c for the T2K experiment at J-PARC. A precise knowledge of differential cross sections for pion and kaon production is of importance for improving the accuracy of neutrino flux simulations. The NA61 detector has a large angular acceptance, full coverage of the T2K phase space region, and good particle identification. In this work the analyses of negatively charged pion production will be presented. Two different methods of negative pion selection and corrections for detector effects will be discussed. Finally, preliminary dn/dp distributions of negatively charged pion in p+C interactions at 31 GeV/c will be presented.

---

[1] Poster presenter.

## 1. Introduction

New experimental information on hadron production in p+C interactions in the region of several tens of GeV is necessary for neutrino beam and cosmic ray event simulations. Even the information on mean multiplicities of the most abundant hadrons is still lacking. It is the region in which the interpolations performed using low energy and high energy models do not agree.

### 1.1 The NA61/SHINE experiment

The physics program of the NA61/SHINE (SHINE = SPS Heavy Ion and Neutrino Experiment) experiment at CERN consists of three subjects: ([1],[2])

1. Measurements of hadron production in hadron-nucleus interactions needed for neutrino (T2K) and cosmic-ray (Pierre Auger and KASCADE) experiments,

2. Measurements of hadron production in proton-proton and proton-nucleus interactions as reference data for nucleus-nucleus reactions,

3. Measurements of energy dependence of hadron production properties in proton-proton, proton-nucleus, and nucleus-nucleus interactions with the aim to identify the properties of the onset of deconfinement and find evidence for the critical point of strongly interacting matter.

### 1.2 The NA61/SHINE detector.

The NA61/SHINE detector is an upgrade of the NA49 experimental setup at CERN [3]. The upgrade mainly concerns the installation of the Forward TOF, a new Time Projection Chamber Read-Out, and a new Data Acquisition System. The Old Zero Degree Calorimeter (ZDC) will be substituted by the Projectile Spectator Detector (PSD). The layout of the NA61/Shine setup and main upgrades are shown in Figure 1.

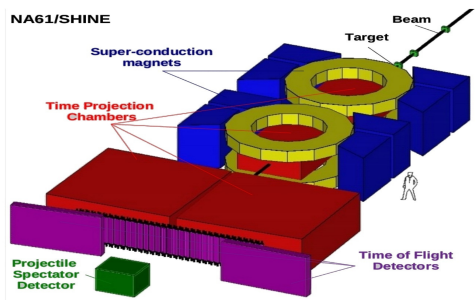

*Figure 1. The layout of the NA61/Shine detector.*

## 2. Measurements of hadron production

In this work three different methods of negatively charged pion analysis will be discussed. The first method of correcting the observed number of tracks, further referred to as h−, is based on the theoretical and experimental premise that the produced negative hadrons at incident energy consists mainly of negative pions with a few percent admixture of negative kaons, electrons from Dalitz decays of neutral mesons, and a negligible fraction of antiprotons. A Monte Carlo VENUS [4] model was used to obtain information on geometric acceptance and reconstruction efficiency, measurement smearing, decays, and non-pion admixture.

The second method was used for a sample of tracks that reached the time of flight detector. Identification of pions was performed using combined dE/dx and time-of-flight information. Factorization of different corrections was assumed and some of them were calculated semi-analytically.

## 3. Results

As a result of h− analysis corrected two dimensional momentum versus emission angle spectrum (Figure 2) and dn/dp distributions in polar angle slices were obtained for negatively charged pions produced from proton interactions.

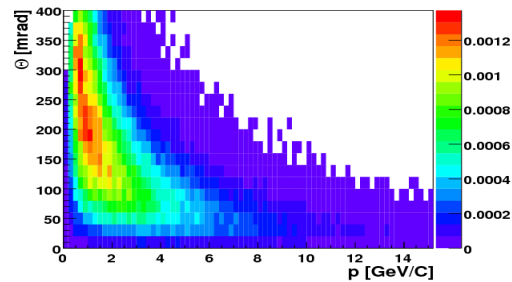

*Figure 2. Corrected spectrum of negatively charged pions produced from proton interactions on a thin carbon target.*

As an example distributions of dn/dp in two selected angular intervals obtained using two presented methods are shown in *Figure 3*

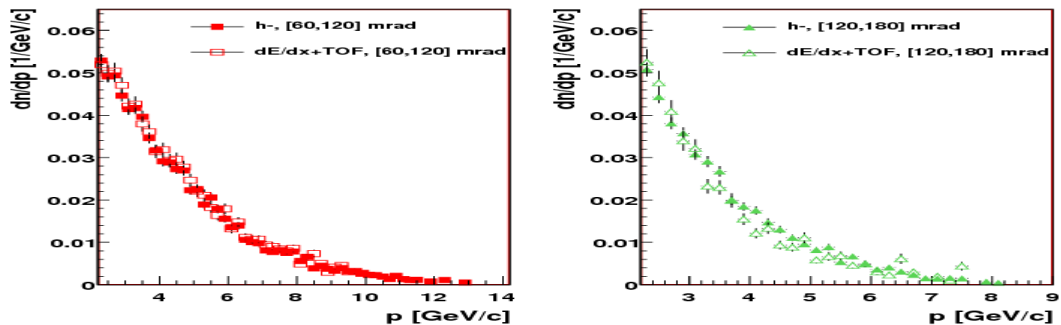

*Figure 3. dn/dp distribution in 60-120 mrad polar angle interval (left) and 120-180 mrad polar angle interval (right) of negatively charged pions from the primary interaction. Results from two different methods are shown.*

The results presented here are preliminary. The work to minimalize systematic biases is in progress. The dependence on the details of the MC generators is also under study.